\documentclass[10pt]{article}

\usepackage{graphicx,amsmath,amsfonts,amssymb,amsthm,lipsum}
\usepackage[font=small]{caption}
\usepackage[labelfont=bf,labelsep=quad,justification=justified]{caption}

\begin{document}

\title{\textbf{From $SL(5,\mathbb{R})$ Yang-Mills theory to induced gravity}}

\author{\textbf{T.~S.~Assimos$^1$}\thanks{assimos@if.uff.br}\ , \textbf{A.~D.~Pereira$^1$}\thanks{aduarte@if.uff.br}\ , \textbf{T.~R.~S.~Santos$^1$}\thanks{tiagoribeiro@if.uff.br}\ ,\\ \textbf{R.~F.~Sobreiro$^1$}\thanks{sobreiro@if.uff.br}\ ,  \textbf{A.~A.~Tomaz$^1$}\thanks{tomaz@if.uff.br}\ ,  \textbf{V.~J.~Vasquez Otoya$^2$}\thanks{victor.vasquez@ifsudestemg.edu.br}\\\\
\textit{{\small $^1$UFF $-$ Universidade Federal Fluminense,}}\\
\textit{{\small Instituto de F\'{\i}sica, Campus da Praia Vermelha,}}\\
\textit{{\small Avenida General Milton Tavares de Souza s/n, 24210-346,}}\\
\textit{{\small Niter\'oi, RJ, Brasil.}}\and
\textit{{\small $^2$IFSEMG $-$ Instituto Federal de Educa\c{c}\~ao, Ci\^encia e Tecnologia,}} \\
\textit{{\small Rua Bernardo Mascarenhas 1283, 36080-001,}}\\
\textit{{\small Juiz de Fora, MG, Brasil}}}
\date{}
\maketitle

\begin{abstract}
From pure Yang-Mills action for the $SL(5,\mathbb{R})$ group in four Euclidean dimensions we obtain a gravity theory in the first order formalism. Besides the Einstein-Hilbert term, the effective gravity has a cosmological constant term, a curvature squared term, a torsion squared term and a matter sector. To obtain such geometrodynamical theory, asymptotic freedom and the Gribov parameter (soft BRST symmetry breaking) are crucial. Particularly, Newton and cosmological constant are related to these parameters and they also run as functions of the energy scale. One-loop computations are performed and the results are interpreted.
\end{abstract}


PACS Number(s): 04.60.--m, 03.50.--z, 04.50.Kd, 11.15.--q
.

\section{Introduction}\label{INTRO}

Four-dimensional metric free gauge theories of gravities have been studied for almost four decades, see for instance \cite{MacDowell:1977jt,Stelle:1979aj,Pagels:1983pq,Gotzes:1989wn,Tresguerres:2008jf,Tseytlin:1981nu,Mielke:2011zza,Sobreiro:2011hb}. This kind of theories are of great relevance on the construction of a quantum gravity model because spacetime deformation and equivalence principle are consequences of the dynamics of the theory, see \cite{Sobreiro:2011hb,Sobreiro:2007pn,Sobreiro:2012dp,Sobreiro:2012iv}. In that sense, the usual incompatibilities between the principles of quantum filed theory and geometrodynamics are avoided. Another important motivation is the fact that all other three fundamental interactions, namely quantum chromodynamics, quantum electrodynamics and weak interactions, are described by gauge theories in flat space \cite{Itzykson:1980rh}. Moreover, quantum field theory can provide reliable predictions through computations in flat space\footnote{For practical reasons, we have to do computations (and therefore, predictions) in \textit{Euclidean} signature spaces, which might be connected to Lorentzian ones by a Wick rotation. It is well-known that, to do concrete computations in standard QFT, a Wick rotation must be performed in order to have a tractable path integral. At the perturbative level, this is nothing else but a change of variables which can always be done. On the other hand, there is no proof that a Wick rotation can be performed at non-perturbative level.}

Mainly, in metric free gauge theories of gravity, the considered gauge groups are de Sitter groups, \emph{i.e.}, $SO(m,n)$ with $m+n=5$ and $m\in\{0,1,2\}$. The main idea is to start with a gauge invariant action and break the group to the Lorentz group in such a way that the $SO(m,n)/SO(m!-1,5-m!)$ sector of the gauge connection is identified with the vierbein and the Lorentz sector with the spin-connection. To do this, a mass scale is necessary because the gauge connection has dimension 1 while the vierbein has dimension 0. Except in \cite{Sobreiro:2011hb,Sobreiro:2007pn}, only topological actions were considered \cite{MacDowell:1977jt,Stelle:1979aj,Pagels:1983pq,Gotzes:1989wn,Tresguerres:2008jf,Tseytlin:1981nu}. In this case, the Higgs mechanism is employed in order to split the group and make the vierbein to emerge. In \cite{MacDowell:1977jt}, however, the breaking mechanism is not discussed and  a dynamical mechanism is not rejected. Examples of alternative gauge groups can be found in \cite{MacDowell:1977jt} for a supergravity description and in \cite{Mielke:2011zza} for a BF theory based on the $SL(5,\mathbb{R})$ group.

Alternative mechanisms to bring geometry from a gauge theory were discussed by some of the authors in \cite{Sobreiro:2011hb,Sobreiro:2007pn,Sobreiro:2012dp,Sobreiro:2012iv}. Specifically, in \cite{Sobreiro:2011hb}, an analogy between quantum chromodynamics and gravity was established. Let us specify the analogy in more detail in the next paragraphs.

Quantum chromodynamics is known to have two distinct phases, the ultraviolet, where quark and gluon states are asymptotically free, and the infrared, where the physical spectrum are hadrons and glueballs (the latter are still theoretical). The final mechanism that connects both sectors still lacks, see for instance \cite{Maas:2011se} and references therein. Among all main proposals to describe such mechanism, there is the Gribov ambiguities problem \cite{Gribov:1977wm,Sobreiro:2005ec} which is necessary to be treated in order to improve the quantization of non-Abelian gauge theories at the low energy regime. Essentially, a residual gauge symmetry survives gauge fixing and this symmetry is relevant only at low energies. Moreover, this problem is inherent to all non-trivial gauge theories and plagues any gauge choice \cite{Singer:1978dk}. Dealing with this problem enforces the introduction of a mass parameter, the so called Gribov parameter \cite{Gribov:1977wm,Zwanziger:1992qr,Dudal:2005na,Dudal:2011gd}, and the breaking of the BRST symmetry in a soft manner \cite{Baulieu:2008fy,Baulieu:2009xr,Dudal:2012sb}. Both effects are responsible for the destabilization of the gluon which acquires complex poles, a feature that is usually interpreted as confinement. Moreover, the fact that the BRST symmetry breaking is a soft breaking allows the usual Yang-Mills theory to be recovered at high energies where the Gribov parameter is very small. Thus, the theory can be continuously deformed into the ultraviolet and infrared sectors, \emph{i.e.}, into the asymptotically free and confined phases. Finally, at the confined phase, due to the gauge principle, one should look for gauge invariant objects in order to determine the physical states of the theory. In the case of quantum chromodynamics, these objects are identified with hadrons and glueballs \cite{Dudal:2010cd,Capri:2012hh,Dudal:2013vha}. It is worth mention that all results obtained are in remarkable agreement with lattice simulations \cite{Cucchieri:2009zt}.

The idea developed in \cite{Sobreiro:2011hb}, and discussed in \cite{Sobreiro:2012dp,Sobreiro:2012iv}, is that gravity could also be described by Yang-Mills theories and soft BRST symmetry breaking. Instead of considering topological theories supplemented with Higgs mechanism, the starting point is the pure Yang-Mills action for de Sitter groups in four-dimensional spacetime. Assuming the existence of a dynamical mass parameter (possibly the Gribov parameter), soft BRST symmetry breaking and asymptotic freedom, the theory suffers an In\"on\"u-Wigner contraction \cite{Inonu:1953sp} which induces a symmetry breaking to the Lorentz group. At this point, the sectors of the gauge field are identified with the vierbein and spin-connection and a first order gravity \cite{Utiyama:1956sy,Kibble:1961ba,Mardones:1990qc} emerges. Thus, at high energies, gravity would be a massless asymptotic free Yang-Mills theory whose spin-1 excitations can be interpreted as the ``gravitons''. As the energy decreases, the soft BRST symmetry takes place and the  mass parameter gradually emerges. The spin-1 ``graviton'' is then ruled out from the spectrum. It is then assumed that, at some lower scale, the ratio $r=\gamma^2/\kappa^2$ ($\gamma$ is the Gribov parameter and $\kappa$ the coupling parameter) is very small and the de Sitter algebra deforms to the Poincar\'e algebra. Since the original action is not invariant under Poincar\'e transformations, the deformation induces a breaking of the gauge group to the Lorentz group. The contraction stage is actually absent in quantum chromodynamics. Finally, evoking the gauge principle, it should be possible to determine the physical observables of the theory. One remarkable feature is that, due to the degrees of freedom of the theory, there are two relatively simple gauge invariant operators that can be identified with nontrivial geometric properties of spacetime, namely, the metric and connection. Thus, instead of composite states, the physical observables are geometrical assignments of spacetime.

Three comments are in order. First, the presence of a mass parameter is crucial because Newton and cosmological constant are not dimensionless. Second, the fact that the mass is dynamical and vanishes at the ultraviolet regime is very important because it prevents gravity to emerge at very high energies. Moreover, the gauge field has dimension 1 while the vierbein has vanishing dimension. Thus, only with a mass at our disposal these quantities could be identified. Third, the mass parameter emerges at the Yang-Mills level due to the presence of Gribov ambiguities in the theory. It means that we must deal with them at the quantum Yang-Mills regime. This is completely different from the discussion of the Gribov problem in gravity theories, as in \cite{Eichhorn:2013ug}. There, spurious configurations of the fluctuation of the metric with respect to a background are present and must be eliminated somehow. In our case, the quantization issues are just present on the Yang-Mills realm.

In the present work we first generalize the method to the $SL(5,\mathbb{R})$ group. Indeed, as aforementioned, this group already has been studied in topological theories, see \cite{Mielke:2011zza}, where the mass parameter responsible by breaking of the $SL(5,\mathbb{R})$ algebra is generated by a Higgs mechanism. Furthermore, it should be clear that any generalization of the $SO(5)$ gauge group should generate, besides of the $SO(4)$ gauge group, an extra matter sector, and this is a motivation to study the $SL(5,\mathbb{R})$. We shall see that, besides cosmological constant, there is an extra sector of the gauge connection which appears as a massive matter field at the final geometrodynamical theory. Second, the quantum sector of the theory is semi-perturbatively analyzed at 1-loop order. Specifically, Newton and cosmological constants are simple functions of the Gribov and coupling parameters (See expression \eqref{eq:newton-cosmol}). Since these quantities are running parameters, the gravitational parameters will run as well. We derive the behaviour of all parameters and show that (for 1-loop approximation) the theory provides very good predictions. More precisely, by fixing the value of Newton's constant, we determine the energy scale of the geometric phase to be of the order of $10^{17}\mathrm{TeV}$, which is right below Planck scale. Moreover, the cosmological constant assumes a very large value which might (at more exact computations) account for the discrepancy between the observed and the quantum vacuum cosmological constants.

This work is organized as follows: In Sect.~\ref{SL5} we provide our definitions and conventions of the $SL(5,\mathbb{R})$ Yang-Mills theory. We also discuss the main properties of the model. In Sect.~\ref{CLASSICAL}, it is shown that the classical action is equivalent to a generalized first order gravity action with a cosmological constant term and a bosonic matter sector. Then, in Sect.~\ref{QUANTUM}, we discuss the validity of this equivalence at quantum level. Specifically, the role of Gribov ambiguities and BRST soft breaking, the running of the parameters and 1-loop estimates are discussed. Finally, our final considerations are displayed in Sect.~\ref{FINAL}.

\section{$SL(5,\mathbb{R})$ gauge theory}\label{SL5}

\subsection{Structure of the $SL(5,\mathbb{R})$ group} \label{group}

The real five-dimensional special linear group, $SL(5,\mathbb{R})$, consists of the collection of all $5\times5$ invertible matrices with unitary determinant. From the $25$ generators of the general linear group $GL(5,\mathbb{R})$, denoted by ${L^A}_B$, we can construct the $24$ trace-free generators of the special linear group, namely
\begin{equation}
{J^{A}}_{B} = {L^{A}}_{B} - \frac{1}{5}{\delta^{A}}_{B}L,\label{gen1}
\end{equation}
where Latin capital indices vary as $\{0,1,2,3,4\}$ and $L={L^A}_A$. Following \cite{Mielke:2011zza}, we choose ${J^{4}}_{4} = 0$. The Killing metric is normalized as $Tr\left({J^{A}}_{B}{J^{C}}_{D}\right) =- {\delta^{A}}_{D}{\delta^{C}}_{B}$. The Lie algebra of the special linear group is given by
\begin{equation}
\left[{J^{A}}_{B}, {J^{C}}_{D}\right]={\delta^{C}}_{B}{J^{A}}_{D}-{\delta^{A}}_{D}{J^{C}}_{B}\;.\label{comutador1}
\end{equation}

According to \cite{Kobayashi:1972book}, there is a four-dimensional matrix representation for the linear group,
\begin{equation}
SL(5,\mathbb{R})\equiv{\mathbb{R}}^{4} \times GL(4, \mathbb{R}) \times {\mathbb{R}}^{4}_{\ast}~,\nonumber
\end{equation}
where the sectors $\mathbb{R}^{4}$ and ${\mathbb{R}}^{4}_{\ast}$ are pseudo-translations. This representation corresponds to a projection on the fourth coordinate $A=(a,4)$. Applying this decomposition on the algebra \eqref{comutador1} and defining $J_{a} = {J^{4}}_{a}$ and $J^{a}_{\ast} = {J^{a}}_{4}$, where small latin indices vary as $\{0,1,2,3\}$, the pseudo-translations algebra is given by
\begin{eqnarray}
\left[J_{a}, J_{b}\right]&=&0\;,\nonumber\\
\left[J^{a}_{\ast}, J^{b}_{\ast}\right]&=&0\;,\nonumber\\
\left[J_{a}, J^{b}_{\ast}\right]&=&-{J^b}_a\;,\label{comutador2}
\end{eqnarray}
while the four-dimensional general linear sector splits as
\begin{eqnarray}
\left[{J^{a}}_{b}, J_{c}\right]&=&-{\delta^{a}}_{c}J_{b}\;,\nonumber\\
\left[{J^{a}}_{b}, J^{c}_{\ast}\right]&=& {\delta^{c}}_{b}J^{a}_{\ast}\;,\nonumber\\
\left[{J^{a}}_{b}, {J^{c}}_d\right]&=&{\delta^{c}}_{b}{J^{a}}_{d}-{\delta^{a}}_{d}{J^{c}}_{b}\;.\label{comutador3}
\end{eqnarray}

We can give a forward step and decompose the $GL(4,\mathbb{R})$ group as $GL(4,\mathbb{R})=SO(4)\times S(10)$, where $S(10)$ denotes a symmetric coset space with $10$ parameters. This decomposition corresponds to write ${J^a}_b={Q^a}_b+{P^a}_b$, where $Q\in SO(4)$ and $P\in S(10)$. Thus, the algebra \eqref{comutador3} will split accordingly. For the mixing with pseudo-translations we have
\begin{eqnarray}
\left[{Q^{a}}_{b}, J_{c}\right]&=&-\frac{1}{2}\left({\delta^{a}}_{c}J_{b}-\delta_{bc}J^a\right)\;,\nonumber\\
\left[{P^{a}}_{b}, J_{c}\right]&=&-\frac{1}{2}\left({\delta^{a}}_{c}J_{b}+\delta_{bc}J^a\right)\;,\nonumber\\
\left[{Q^{a}}_{b}, J^{c}_{\ast}\right]&=&\frac{1}{2}\left({\delta^{c}}_{b}J^{a}_{\ast}-\delta^{ac}J_{\ast b}\right)\;,\nonumber\\
\left[{P^{a}}_{b}, J^{c}_{\ast}\right]&=&\frac{1}{2}\left({\delta^{c}}_{b}J^{a}_{\ast}+\delta^{ac}J_{\ast b}\right)\;,\label{comutador4}
\end{eqnarray}
and for the $GL(4,\mathbb{R})$ sector the decomposition reads
\begin{eqnarray}
\left[{Q^{a}}_{b}, {Q^{c}}_d\right]&=&-\frac{1}{2}\left({\delta^a}_d{Q^c}_b - \delta_{bd}Q^{ca} + \delta^{ca}Q_{bd}-{\delta^c}_b{Q^a}_d\right)\;,\nonumber\\
\left[{P^{a}}_{b}, {P^{c}}_d\right]&=&\frac{1}{2}\left({\delta^c}_b{Q^a}_d+ \delta^{ac}Q_{bd}-\delta_{bd}Q^{ca}-{\delta^a}_d{Q^c}_b\right)\;,\nonumber\\
\left[{Q^{a}}_{b}, {P^{c}}_d\right]&=&\frac{1}{2}\left({\delta^c}_b{P^a}_d - \delta^{ac}P_{bd} + \delta_{bd}P^{ac}-{\delta^a}_d{P^c}_b\right)\;.\label{comutador5}
\end{eqnarray}
Thus, after the entire decomposition
\begin{equation}
SL(5,\mathbb{R})\equiv{\mathbb{R}}^{4}\times\left[SO(4) \times S(10)\right]\times{\mathbb{R}}^{4}_{\ast}~,\nonumber
\end{equation}
the relevant algebra is \eqref{comutador2}, \eqref{comutador4} and \eqref{comutador5}.

A few remarks are in order: i.) The group $SO(4)$ is a stability group with respect to both groups, namely, $SL(5,\mathbb{R})$ and $GL(4,\mathbb{R})$. This is evident from the algebra decompositions \eqref{comutador4} and \eqref{comutador5}. ii.) The symmetric space $S(10)$ is trivial in the sense that it can be contracted down to a point \cite{Kobayashi:1963book}. iii.) The pseudo-translations are also symmetric spaces. However, due to the third relation in \eqref{comutador2}, this is not a trivial space. iv.) The pseudo-translations may become Abelian through an In\"on\"u-Wigner contraction (Physically, this contraction requires a mass scale) \cite{Inonu:1953sp} and then, eventually, becoming a trivial space.

\subsection{Fields and action}\label{gauge}

The starting point of this work is an $SL(5,\mathbb{R})$ Yang-Mills massless action in a Euclidean four-dimensional spacetime,
\begin{equation}
S_{\mathrm{YM}}=\int{F_A}^B\ast{F_B}^A\;,\label{ym1}
\end{equation}
where ${F^A}_B$ are the components of the field strength 2-form $F = dY + \kappa YY$, $d$ is the exterior derivative, $\kappa$ is the coupling parameter and $Y=dx^\mu{Y_{A\mu}}^B{J^A}_B$ is the algebra valued gauge connection 1-form. The action \eqref{ym1} is invariant under $SL(5,\mathbb{R})$ gauge transformations, \emph{i.e.}, $Y\longmapsto U^{-1}\left(\frac{1}{\kappa}d+Y\right)U$, where $U\in SL(5,\mathbb{R})$. At infinitesimal level, the gauge transformations reduce to
\begin{equation}
Y\longmapsto Y+\mathcal{D}\alpha\;,\label{gt1}
\end{equation}
where $\mathcal{D}=d+\kappa Y$ is the adjoint covariant derivative. 

Spacetime and the gauge group are not dynamically related. Moreover, since this is a massless theory, there cannot be any relation between the degrees of freedom of this action with gravity. This occurs because the gauge field has UV dimension 1 and the vierbein (or the metric tensor) has vanishing dimension. Moreover, a mass parameter is required to make Newton's constant to emerge. Before we discuss the masses that can be dynamically generated, we apply the group decomposition discussed in Sect.~\ref{group} to the action \eqref{ym1}. The connection splits as
\begin{equation}
Y={Y_A}^B {J^A}_B = {A_a}^b{Q^a}_b+{M_a}^b{P^a}_b+\theta^aJ_a+\pi_aJ_*^a\;,\label{connec1}
\end{equation}
and the corresponding field strength also decomposes,
\begin{eqnarray}
F={F_A}^B{J^A}_B&=&\left[{F_a}^b+\kappa{M_a}^c{M_c}^b+\frac{\kappa}{2}\left(\pi_a\theta^b-\pi^b\theta_a\right)\right]{Q^a}_b + \left[\nabla {M_a}^b+\frac{\kappa}{2}\left(\pi_a\theta^b+\pi^b\theta_a\right)\right]{P^a}_b \nonumber\\
&+&\left(\nabla\theta^a-\kappa{M_b}^a\theta^b\right)J_a + \left(\nabla\pi_{a}+\kappa{M_a}^b\pi_b\right)J_{*}^{a}\;,\nonumber\\
\end{eqnarray}
where, $\nabla = d + \kappa A$ denotes the covariant derivative with respect to the $SO(4)$ sector and ${F_a}^b=d{A_a}^b+\kappa{A_a}^c{A_c}^b$. Thus, the action \eqref{ym1} decomposes as
\begin{eqnarray}
 S_{\mathrm{YM}}& = & \int\left\{{F_a}^b\ast{F_b}^a+2\kappa{F_a}^b\ast \left(\pi_b\theta^a\right) + 2\nabla\theta^a\ast\nabla\pi_a + \kappa^2(\pi_a\theta^b)\ast\left(\pi_b\theta^a\right) \right.\nonumber\\
 &+&\left. \nabla {M_a}^b\ast\nabla {M^a}_b + \kappa^2{M_a}^c{M_c}^b\ast \left({M_b}^d{M_d}^a\right) + 2\kappa{F_a}^b\ast\left({M_b}^c{M_c}^a\right)+ 2\kappa\nabla\theta^a\ast\left({M_a}^b\pi_b\right)\right.\nonumber\\
 &-&\left. 2\kappa\nabla\pi_a\ast\left({M^a}_b\theta^b\right) + 2\kappa\nabla{M_a}^b\ast\left(\pi_b\theta^a\right) - 2\kappa^2{M^a}_b\theta^b\ast\left({M_a}^c\pi_c\right) + 2\kappa^2{M_a}^c{M_c}^b \ast\left(\pi_b\theta^a\right)\right\}\;\nonumber\\.\label{ym2}
 \end{eqnarray}

The gauge transformations \eqref{gt1} are then separated into sectors as
\begin{eqnarray}
{A_a}^b \longmapsto {A_a}^b &+& d{\zeta_a}^b + \kappa\left({A_a}^c{\zeta_c}^b-{A^b}_c{\zeta^c}_a + {M_a}^c{\xi_c}^b-{M^b}_c{\xi^c}_a\right) - \frac{\kappa}{2}\left(\theta_a\beta^b- \theta^b\beta_a-\pi_a\eta^b+\pi^b\eta_a\right)\;,\nonumber\\
{M_a}^b \longmapsto {M_a}^b &+& d{\xi_a}^b + \kappa\left({M_a}^c{\zeta_c}^b-{M^b}_c{\zeta_a}^c + {A_a}^c{\xi_c}^b+{A^b}_c{\xi^c}_a\right) - \frac{\kappa}{2}\left(\theta_a\beta^b+\theta^b\beta_a-\pi_a\eta^b-\pi^b\eta_a\right)\;,\nonumber\\
\theta^a \longmapsto \theta^a &+& d\eta^a + \kappa\left(A^a_{\ b}\eta^b-M^a_{\ b}\eta^b-\theta^b\zeta^a_{\ b}+\theta^b\xi^a_{\ b}\right)\;,\nonumber\\
\pi_a \longmapsto \pi_a &+& d\beta_a + \kappa\left(A_a^{\ b}\beta_b+M_a^{\ b}\beta_b-\pi_b\zeta_a^{\ b}+\pi_b\xi_a^{\ b}\right)\;,\nonumber\\\label{gt3}
\end{eqnarray}
where the gauge parameter was dismembered as
\begin{equation}
\alpha= {\zeta_a}^b{Q^a}_b + {\xi_a}^b{P^a}_b + \eta_aJ^a + \beta_aJ^a_*~.\nonumber
\end{equation}

For further use, we remark that the action \eqref{ym2} is invariant under the discrete symmetry
\begin{eqnarray}
A&\longmapsto&A\;,\nonumber\\
M&\longmapsto&-M\;,\nonumber\\
\theta&\longmapsto&\pi\;,\nonumber\\
\pi&\longmapsto&\theta\;.\label{sym}
\end{eqnarray}
Essentially, this symmetry establishes that fields $\pi$ and $\theta$ are indistinguishable.

\subsection{Physical properties}\label{phys}

The action \eqref{ym2} is nothing else than the $SL(5,\mathbb{R})$ Yang-Mills action \eqref{ym1} in a decomposed form. The Yang-Mills theory has two main properties \cite{Itzykson:1980rh}:

\begin{itemize}

\item \emph{Renormalizability}: The Yang-Mills action defines a renormalizable theory, at least to all orders in perturbation theory \cite{Piguet:1995er}. Renormalizability states that ultraviolet divergences can be consistently eliminated from perturbative computations, \emph{i.e.}, Yang-Mills theories are stable at quantum level. It is worth mention that BRST symmetry is very important with respect to the renormalizability of a gauge theory.

\item \emph{Asymptotic freedom}: The renormalization of the coupling parameter leads to the concept of asymptotic freedom \cite{Gross:1973id,Politzer:1973fx}. This property predicts that, at high energies, the coupling is very small and perturbation theory can be safely employed. However, as the energy decreases, the coupling increases. At this strong coupling regime, the theory is non-perturbative and yet to be fully understood.

\end{itemize}

It was shown in \cite{Gribov:1977wm} that, at low energies, gauge fixing is not possible by the simple introduction of a constraint. This problem is commonly known as Gribov ambiguities. The main point in the elimination of the Gribov ambiguities is that BRST quantization (and also the standard Faddeev-Popov quantization) shows itself to be incomplete at low energy level. In fact, the implementation of a gauge fixing does not eliminate completely the gauge symmetry, a residual symmetry survives. Moreover, it was shown that this problem occurrs for any gauge choice \cite{Singer:1978dk}. The spurious gauge configurations are called Gribov copies and they are the kernel of the Gribov ambiguities. Remarkably, these copies gain relevance only at low energies, keeping the high energy sector untouched and consistent at quantum level. To eliminate the Gribov ambiguities is a hard step that is not yet fully understood. However, at the Landau gauge, the infinitesimal copies can be eliminated through the introduction of a soft BRST breaking related to a mass parameter known as Gribov parameter $\gamma$. It is also remarkable that, in QCD, the treatment of this technical problem leads to exceptional evidences of quark-gluon confinement \cite{Gribov:1977wm,Zwanziger:1992qr,Maggiore:1993wq,Dudal:2005na,Dudal:2011gd}. The main properties of dealing with Gribov ambiguities are the following:

\begin{itemize}

\item The elimination of infinitesimal Gribov ambiguities at the Landau gauge leads to the so called Gribov-Zwanziger local action \cite{Zwanziger:1992qr,Maggiore:1993wq,Dudal:2005na} which carries extra auxiliary fields associated to a non-local term that accounts for the infinitesimal Gribov copies. This non-local term is proportional to $\gamma^4$. Remarkably, at the ultraviolet limit, this term is suppressed and the usual Yang-Mills action is recovered. The Gribov-Zwanziger action is renormalizable, at least to all orders in perturbation theory, in such a way that the ultraviolet sector of the theory is unchanged. In fact, no extra renormalizations are required \cite{Zwanziger:1992qr,Dudal:2005na}; all extra fields and the Gribov parameter renormalization factors depend on the gluon and coupling parameter renormalization factors.

\item It turns out that the Gribov-Zwanziger action is not BRST invariant, and the breaking is proportional to the Gribov parameter squared. Since $\gamma$ has dimension of a mass, the breaking is soft, and thus, harmless to the ultraviolet sector \cite{Baulieu:2008fy,Baulieu:2009xr}. It was shown that, the breaking can be controlled and does not affect the quantum sector of the model.

\item The Gribov parameter is determined from the minimization of the quantum action $\delta\Sigma/\delta\gamma^2=0$. A straightforward computation at tree level \cite{Gribov:1977wm} predicts
\begin{equation}
\gamma^2=\mu^2 \exp{\left(\frac{-64\pi^2}{3N\kappa^2}\right)}~,\nonumber
\end{equation}
where $\mu^2$ is a cutoff and $N$ is the Casimir of the gauge group, $f^\mathcal{ABC}f^\mathcal{BCD}=-N\delta^\mathcal{AB}$, where the collective index means $\mathcal{A}\equiv\{AB\}$. Thus, at the perturbative regime, $\kappa\rightarrow0$, we have that $\gamma\rightarrow0$. To take this limit is equivalent to take the high energy limit. Thus, the presence of the Gribov parameter allows the Gribov-Zwanziger action to be continuously deformed into the Yang-Mills action. In fact, the ultraviolet and infrared sectors can be continuously deformed one in each other \cite{Baulieu:2008fy,Baulieu:2009xr}.

\item Besides the Gribov parameter, another possible effect is that, at low energies, dynamical mass generation \cite{Dudal:2005na,Dudal:2011gd} takes place. It originates from the condensation of dimension-2 operators. The combination of these mass parameters and the Gribov parameter drastically changes the behavior of the theory at infrared scale. In the case of $SU(N)$ group, these changes are also relevant for confinement.

\item One of the evidences of confinement emerges from the gauge propagator which, for finite values of $\gamma$, acquires complex poles \cite{Dudal:2005na,Dudal:2011gd}. The consequence is that it violates positivity of the spectral representation and thus, no physical particles can be associated with this propagator.

\item The union of the Gribov-Zwanziger formalism and the condensation of dimension-2 operators is called the refined Gribov-Zwanziger formalism \cite{Dudal:2011gd} and improves the results obtained from the Gribov-Zwanziger action. For instance, the propagators of the RGZ action coincide with the results obtained from huge lattice simulations \cite{Cucchieri:2009zt}.

\item We also remark that, some advances on the determination of the physical spectrum of the infrared sector of Yang-Mills theories have been made \cite{Dudal:2010cd,Capri:2011ki}. Although very difficult, the main requirement is gauge invariance, \emph{i.e.}, observables are identified with gauge invariant operators which have spectral representation. In quantum chromodynamics, these operators must describe its low energy spectrum, \emph{i.e.}, hadrons and glueballs.
\end{itemize}

\section{Effective gravity}\label{CLASSICAL}

We will now show how the action \eqref{ym2} can be associated with a geometrodynamical gravity theory. At this point we consider only the action \eqref{ym2} and the existence of a mass parameter $\gamma$ (possibly, the Gribov parameter). This is justified by the fact that the quantum action $\Sigma$ is supposed to be of the same form of the classical action \cite{Piguet:1995er}. The effect depends on the existence of such a mass that tends to zero at the ultraviolet regime. At Sect.~\ref{QUANTUM}, we will discuss some properties and consequences of quantum effects on the light of Gribov ambiguities and BRST soft breaking.

\subsection{Rescalings and contractions}

The first step is a rescaling of the fields, only possible if a mass scale is at our disposal. We perform the following rescaling of the fields
\begin{eqnarray}
\{{A_a}^b,{M_a}^b\}&\mapsto&\kappa^{-1}\{{A_a}^b,{M_a}^b\}\;,\nonumber\\
\{\theta^a,\pi_a\}&\mapsto&\gamma\kappa^{-1}\{\theta^a,\pi_a\}\;.\label{resc1}
\end{eqnarray}
Under this rescaling, the action \eqref{ym2} now reads
\begin{eqnarray}
 S_{\mathrm{YM}}&=&\frac{1}{\kappa^2}\int\left\{ {\overline{F}_a}^b \ast{\overline{F}_b}^a + 2\gamma^2{\overline{F}_a}^b\ast\left(\pi_b\theta^a\right) +  2\gamma^2\overline{\nabla}\pi^a\ast\overline{\nabla}\theta_a + \gamma^4\pi_a\theta^b\ast\left(\pi_b\theta^a\right) + \overline{\nabla}{M_a}^b\ast\overline{\nabla}{M^a}_b \right.\nonumber\\
&+&\left. {M_a}^c{M_c}^b\ast\left({M_b}^d{M_d}^a\right) + 2{\overline{F}_a}^b\ast \left({M_b}^c{M_c}^a\right) + 2\gamma^2\overline{\nabla}\theta^a\ast\left({M_a}^b \pi_b\right) - 2\gamma^2\overline{\nabla}\pi^a\ast\left({M_a}^b \theta_b\right) \right.\nonumber\\
 &+&\left. 2\gamma^2\overline{\nabla}{M_a}^b\ast\left(\pi_b\theta^a\right) - 2\gamma^2{M^a}_b\theta^b\ast\left({M_a}^c\pi_c\right)+ 2\gamma^2{M_a}^c{M_c}^b\ast\left(\pi_b\theta^a\right) \right\}\;.\nonumber\\\label{ym3}
 \end{eqnarray}
We remark that, $\theta^a$ and $\pi^a$ now have vanishing dimension and thus, are suitable candidates to be associated with the vierbein. On the other hand, $A$ and $M$ remain with dimension-1. Moreover, the common factor $\kappa^{-1}$ in the rescaling is a standard procedure in quantum field theory \cite{Itzykson:1980rh}. The effect is that it is completely factorized at the action. This is evident from the expressions of the over-lined quantities, $\overline{F}=d{A_a}^b+{A_a}^c{A_c}^b$ and $\overline{\nabla}= d + A$.

In order to preserve the structure of the algebra-valued fields $\theta$ and $\pi$, the mapping has to be imposed for the algebras as well, by $\{J_a,J^a_\ast\}\longmapsto \kappa\gamma^{-1}\{J_a,J^a_\ast\}$, see \cite{Sobreiro:2011hb}. This extra requirement ensures that, under the rescaling, $\{\theta,\pi\}\longmapsto \{\theta,\pi\}$. This affects only the third relation of \eqref{comutador2} by
\begin{equation}
\left[J_a,J^b_\ast\right]=-\frac{\gamma^2}{\kappa^2}{J^b}_a\;.\label{comutador6}
\end{equation}

The next step consists on the deformation of the theory based on the group $SL(5,\mathbb{R})$ into a reduced theory with $SO(4)$ gauge invariance. We can make this contraction in two distinct steps. First, at low energy regime, the presence of a mass parameter allowed the rescaling of the fields that led to the action \eqref{ym3}. The consequence for the algebra is described at the relation \eqref{comutador6}. Thus, for a regime in which $\gamma^2/\kappa^2\rightarrow 0$, the pseudo-translations become two independent pairs of translations. This is ensured by the In\"on\"u-Wigner theorem \cite{Inonu:1953sp}. Thus, from the triviality of translations, the group can be continuously deformed as $SL(5,\mathbb{R})\longmapsto GL(4,\mathbb{R})$, see for instance \cite{Kobayashi:1963book,Sobreiro:2011hb}. Second, the contraction $GL(4,\mathbb{R})\longmapsto SO(4)$ is also ensured because the sector $S(10)$ is trivial (It is isomorphic to a vector space \cite{Kobayashi:1963book}). Thus, it is a standard theorem on fiber bundle theory \cite{Kobayashi:1963book}, that the reduced connection $A$ defines a connection on the $SO(4)$ gauge theory. The rest of the fields survive as matter fields. In fact, after all contractions, the gauge transformations \eqref{gt3} reduce to
\begin{eqnarray}
{A_a}^b&\longmapsto&{A_a}^b+\overline{\nabla}{\zeta_a}^b\;,\nonumber\\
{M_a}^b&\longmapsto&{M_a}^b+{M_a}^c{\zeta_c}^b-{M^b}_c{\zeta_a}^c\;,\nonumber\\
\theta^a&\longmapsto&\theta^a-\theta^b\zeta^a_{\ b}\;,\nonumber\\
\pi_a&\longmapsto&\pi_a-\pi_b\zeta_a^{\ b}\;,\label{gtc}
\end{eqnarray}
which is the typical gauge transformations for the $SO(4)$ gauge theory. It is clear that $A$ is the gauge field and the rest are matter fields.

\subsection{Observables, geometry and gravity}

We are now ready to identify the $SO(4)$ gauge theory described by the action \eqref{ym3} with gravity. To do this, it is crucial to identify a few gauge invariant operators. The question that should be addressed is: \emph{What is the IR theory we want and which are the observables?} In electroweak theory, the infrared sector suffers a spontaneous symmetry breaking and the theory has to be redefined in order to obtain its correct physical spectrum, described by gauge invariant operators. In quantum chromodynamics, although a final theoretical description is yet to be developed, it is known that the gauge invariant operators must be associated with hadrons and glueballs. In gravity, inevitably, these operators have to be identified with geometry. In fact, it is easy to see that the two operators
\begin{eqnarray}
\sigma_{\mu\nu}&=&\delta_{ab}\theta^a_\mu\pi^b_\nu\;\nonumber\\
{\Theta^\alpha}_{\mu\nu}&=&\delta_{ab}\sigma^{\alpha\beta}\theta^b_\beta\left(\partial_\mu\pi^a_\nu+{A^a}_{\mu\ c}\pi^c_\nu\right)\;,\label{gi1}
\end{eqnarray}
are gauge invariant and carry the degrees of freedom of the metric tensor $g_{\mu\nu}$ and affine connection ${\Gamma^\alpha}_{\mu\nu}$. Thus, one can identify an effective geometry from
\begin{eqnarray}
g_{\mu\nu}&=&\left<\sigma_{\mu\nu}\right>\;,\nonumber\\
{\Gamma^\alpha}_{\mu\nu}&=&\left<{\Theta^\alpha}_{\mu\nu}\right>\;,\label{geom1}
\end{eqnarray} 
where the expectation value should be taken with respect to the most complete action (see next section). The relations \eqref{geom1} can actually be performed by identifying the fields $\pi$ and $\theta$ with the vierbein of the effective spacetime and $A$ with the spin-connection. The symmetry between $\theta$ and $\pi$ described in \eqref{sym} is crucial, otherwise, it would not be possible to identify both translational fields with the vierbein. Thus, under this assumption, the gauge theory constructed in $\mathbb{R}^4$ can be identified with a deformed spacetime $\mathbb{M}^4$. To make this identification consistent, we follow \cite{Sobreiro:2011hb}: First, we impose that p-forms in $\mathbb{R}^4$  are mapped into p-forms in $\mathbb{M}^4$ and, as a consequence, Hodge duals in $\mathbb{R}^4$ are mapped into Hodge duals in $\mathbb{M}^4$. See \cite{Sobreiro:2011hb,Sobreiro:2012iv} for more formal discussions on this mapping. In addition, after the breaking, the fields $\theta$ and $\pi$ are redundant. This feature and the symmetry \eqref{sym} allow both fields to be identified with the vierbein. Thus, the following consistent identifications are demanded
\begin{eqnarray}
{\omega_\mathfrak{a}}^\mathfrak{b}&=&\delta_\mathfrak{a}^a \delta_b^\mathfrak{b} {A_a}^b\;,\nonumber\\
\frac{1}{2}e^\mathfrak{a}&=&\delta^\mathfrak{a}_{a}\theta^a\;\;=\;\;\delta^\mathfrak{a}_a\pi^a\;,\nonumber\\
{m_\mathfrak{a}}^\mathfrak{b}&=&\delta_\mathfrak{a}^a \delta_b^\mathfrak{b} {M_a}^b\;,\label{geom2}
\end{eqnarray}
where $\omega$ is the spin-connection and $m$ a symmetric 1-form. Latin indices $\mathfrak{a,b,c..}$ refer to the tangent space $T_X(\mathbb{M})$ in $X\in\mathbb{M}^4$.

The map \eqref{geom2}, applied to the action \eqref{ym3}, yields
\begin{equation}
S_{\mathrm{YM}}\;\longmapsto\; S_{map}~,\nonumber
\end{equation}
which can be explicitly written as

\begin{eqnarray} \label{ym-map}
S_{map}&=&\frac{\gamma^2}{\kappa^2}\int \left\{\frac{1}{\gamma^2}{R_\mathfrak{a}}^\mathfrak{b}\star {R_\mathfrak{b}}^\mathfrak{a}+\frac{1}{2}{R_\mathfrak{ab}} \star\left(e^\mathfrak{b}e^\mathfrak{a}\right) + \frac{1}{2}T^\mathfrak{b}\star T_\mathfrak{b} + \frac{\gamma^2}{16}e_\mathfrak{a}e_\mathfrak{b}\star \left(e^\mathfrak{b}e^\mathfrak{a}\right) + \frac{1}{\gamma^2}D {m_\mathfrak{a}}^\mathfrak{b}\star D {m_\mathfrak{b}}^\mathfrak{a} \right.\nonumber\\
&+&\left.  \frac{1}{\gamma^2}{m_\mathfrak{a}}^\mathfrak{c}{m_\mathfrak{c}}^\mathfrak{b} \star ({m_\mathfrak{b}}^\mathfrak{d}{m_\mathfrak{d}}^\mathfrak{a}) + \frac{2}{\gamma^2}{R_\mathfrak{a}}^\mathfrak{b}\star({m_\mathfrak{b}}^\mathfrak{c}{m^\mathfrak{a}}_\mathfrak{c}) - \frac{1}{2}{m_\mathfrak{b}}^\mathfrak{a} e^\mathfrak{b}\star ({m_\mathfrak{a}}^\mathfrak{c} e_\mathfrak{c})+ \frac{1}{2} m_\mathfrak{al}{m^\mathfrak{l}}_\mathfrak{b} \star\left(e^\mathfrak{b}e^\mathfrak{a}\right) \right\}\;.
\end{eqnarray}
This action can be interpreted as a four-dimensional gravity if we identify Newton and cosmological constants through
\begin{equation} \label{eq:newton-cosmol}
\gamma^2=\frac{\kappa^2}{8\pi G}=\frac{4\Lambda^2}{3}\;,
\end{equation}
where $G$ is the Newton constant and $\Lambda^2$ stands for the cosmological constant. Thus, we have for the effective gravity theory
\begin{eqnarray}\label{ym-map-grav}
 S_{\mathrm{Grav}} &=& \frac{1}{16\pi G}\int \left\{-\frac{3}{2\Lambda^2}{R_\mathfrak{a}}^\mathfrak{b}\star {R^\mathfrak{a}}_\mathfrak{b} -\frac{1}{2}\epsilon_\mathfrak{abcd}{R}^\mathfrak{ab} e^\mathfrak{c}e^\mathfrak{d} + T^\mathfrak{b}\star T_\mathfrak{b} - \frac{\Lambda^2}{12}\epsilon_\mathfrak{abcd}e^\mathfrak{a}e^\mathfrak{b}e^\mathfrak{c}e^\mathfrak{d}+ \frac{3}{2\Lambda^2}D {m_\mathfrak{a}}^\mathfrak{b}\star D {m_\mathfrak{b}}^\mathfrak{a} \right.\nonumber\\
 &+&\left. \frac{3}{2\Lambda^2}{m_\mathfrak{a}}^\mathfrak{c}{m_\mathfrak{c}}^\mathfrak{b} \star ({m_\mathfrak{b}}^\mathfrak{d}{m_\mathfrak{d}}^\mathfrak{a})+ \frac{3}{\Lambda^2}{R_\mathfrak{a}}^\mathfrak{b}\star ({m_\mathfrak{b}}^\mathfrak{c} {m^\mathfrak{a}}_\mathfrak{c}) - {m_\mathfrak{b}}^\mathfrak{a} e^\mathfrak{b}\star ({m_\mathfrak{a}}^\mathfrak{c} e_\mathfrak{c}) -\frac{1}{2}\epsilon^\mathfrak{abcd} m_\mathfrak{al}{m^\mathfrak{l}}_\mathfrak{b} e_\mathfrak{c} e_\mathfrak{d} \right\}.\nonumber\\
\end{eqnarray}
The action \eqref{ym-map-grav} describes a first order gravity coupled to a matter field $m$. The Einstein-Hilbert term, as well the cosmological constant, are immediately recognized. Moreover, quadratic terms on the curvature and torsion are present, providing a modification of standard general relativity. The Newton constant is expected to be very weak at the present stage of the Universe. Thus, from \eqref{eq:newton-cosmol}, the cosmological constant should be very large. This is a good feature of the model because this pure (renormalized) gravitational constant $\Lambda_{ren}^2$ can enter in the game together with the cosmological constant predicted by quantum field theory. Thus, there is hope in order to obtain a complete cosmological constant that coincides with the observations \cite{Weinberg:1988cp,Shapiro:2009dh}. In fact, a compensation can occur by imposing a renormalization condition for the model:
\begin{equation}
\Lambda_{obs}^2=\Lambda_{ren}^2+\Lambda_{qft}^2\;.\label{lambdas}
\end{equation}

Concerning about Eq.~\eqref{lambdas}, the main idea is to fix the Newton's constant, which depends on $\kappa$ and $\gamma$, as its experimental value. Once we fix it, the renormalization group cutoff is automatically set near to the Planck scale. Hence, the cosmological constant inherent to the model is fixed to a huge value -- See details in Sect.~\ref{QUANTUM}. We still have to consider the vaccum of the matter content, eventually. So, since the matter content provides a huge, but negative, cosmological constant, we argue that the sum of this value with ours (positive huge value) could provide the correct value for an observational cosmological constant. The main idea about this point is discussed in detail in reference \cite{Shapiro:2009dh}.

Concerning the matter sector, it is clear that $m$ is massive (last two terms in \eqref{ym-map-grav}), and thus, has a finite range. Moreover, due to the symmetry of its tangent indices, it carries up to spin-3 modes. The origin of such ``matter sector" is the decomposition of what would be the non-metricity of the nontrivial geometry. Such exchange among geometric properties of the effective geometry and matter fields was explored in \cite{Sobreiro:2010ji}. The very identification of such matter fields is not enough to attach a physical meaning to them and a detailed analysis is necessary. 

For completeness, we write the field equations from action \eqref{ym-map-grav}. For the vierbein, spin-connection and the matter field, we obtain, respectively,
\begin{eqnarray} \label{eq:eqfield-e}
 &-& \frac{3}{2\Lambda^2}R^{\mathfrak{bc}} \star (R_{\mathfrak{bc}}e_{\mathfrak{a}}) + D\star T_\mathfrak{a} + T^\mathfrak{b}\star(T_\mathfrak{b} e_\mathfrak{a})-\epsilon_\mathfrak{abcd}{R}^\mathfrak{bc}e^\mathfrak{d}
 -\frac{\Lambda^2}{3}\epsilon_\mathfrak{abcd}e^\mathfrak{b}e^\mathfrak{c}e^\mathfrak{d}= -\frac{3}{2\Lambda^2} \left[D m^\mathfrak{b} \,_\mathfrak{c}\star (D m_\mathfrak{b} \,^\mathfrak{c} e_\mathfrak{a})\right.\nonumber\\
&+&\left. m_\mathfrak{b} \,^\mathfrak{c}m_\mathfrak{c} \,^\mathfrak{d} \star (m_\mathfrak{d} \,^\mathfrak{l}m_\mathfrak{l} \,^\mathfrak{b}e_\mathfrak{a})
+2{R}^\mathfrak{b}\,_\mathfrak{c}\star ({m}^\mathfrak{c}\,_\mathfrak{d} m^\mathfrak{db} e_\mathfrak{a}) \right]+  m^\mathfrak{b}\,_\mathfrak{a}\star (m^\mathfrak{c}\,_\mathfrak{b} e_\mathfrak{c})-m_\mathfrak{b}\,^\mathfrak{c} e_\mathfrak{c}\star (m_\mathfrak{d}\,^\mathfrak{b} e^\mathfrak{d}e_\mathfrak{a})
+\epsilon_\mathfrak{abcd} m^\mathfrak{b}\,_\mathfrak{l} m^\mathfrak{lc} e^\mathfrak{d} \;,\nonumber\\
\end{eqnarray}
\begin{eqnarray} \label{eq:eqfield-o}
& &\frac{3}{\Lambda^2}D\star R_\mathfrak{ab}+ \epsilon_\mathfrak{abcd}T^\mathfrak{c}e^\mathfrak{d}
- e_\mathfrak{a}\star T_\mathfrak{b}+e_\mathfrak{b}\star T_\mathfrak{a}= -\frac{3}{\Lambda^2} \left[m_\mathfrak{a}\;^\mathfrak{c} \star D m_{\mathfrak{c}\mathfrak{b}}-m_\mathfrak{b}\;^\mathfrak{c} \star D m_{\mathfrak{c}\mathfrak{a}}+D\star(m_\mathfrak{a}\;^\mathfrak{c} m_{\mathfrak{c}\mathfrak{b}})\right],\nonumber\\
\end{eqnarray}
\begin{eqnarray}\label{eq:eqfield-m}
&\ &D\star D m_\mathfrak{ab} + m_\mathfrak{a}\;^\mathfrak{c}\star(m_\mathfrak{c}\;^\mathfrak{d} m_\mathfrak{db})+  m_\mathfrak{b}\;^\mathfrak{c}\star(m_\mathfrak{c}\;^\mathfrak{d} m_\mathfrak{da})+m_\mathfrak{a}\;^\mathfrak{c}\star R_\mathfrak{cb}+m_\mathfrak{b}\;^\mathfrak{c}\star R_\mathfrak{ca} \nonumber\\
&-& \frac{\Lambda^2}{3}\left[e_\mathfrak{b}\star(m_\mathfrak{ac}e^\mathfrak{c}) +
e_\mathfrak{a}\star(m_\mathfrak{bc}e^\mathfrak{c})\right] -\frac{\Lambda^2}{3}\left(\epsilon_\mathfrak{acdl}m^\mathfrak{l}\;_\mathfrak{b}e^\mathfrak{c}e^\mathfrak{d}+
\epsilon_\mathfrak{bcdl}m^\mathfrak{l}\;_\mathfrak{a}e^\mathfrak{c}e^\mathfrak{d}\right)= 0\;.
\end{eqnarray}
%\end{widetext}

\section{Quantum aspects, consistency checks and further improvements}\label{QUANTUM}

To show that the classical $SL(5,\mathbb{R})$ Yang-Mills action is equivalent to the gravity theory \eqref{ym-map-grav}, there were made three hypothesis: First, that a mass emerges at some scale, otherwise, no vierbein can be defined. Second, that the ratio $\gamma^2/\kappa^2$ is sufficient small at some scale. Third, that there exists a map between the original Euclidean base space and the effective deformed spacetime. Let us discuss each of these issues and other relevant details. For simplicity, we will restrict ourselves to the Landau gauge and we will fix our attention exclusively to the Gribov mass.

\subsection{Gribov ambiguities and soft BRST symmetry breaking}

The model needs a mass parameter that could be used to separate two sectors of the theory: the quantum sector (massless perturbative Yang-Mills theory) and the effective sector (geometrodynamical gravity theory). The most natural mass parameter is the Gribov parameter. Let us identify it with $\gamma$. This parameter appears as a necessity in order to keep quantum consistency of Yang-Mills theories at low energy regime \cite{Gribov:1977wm,Zwanziger:1992qr,Dudal:2005na} and enters in the action through a non-local extra term. For instance, at the Landau gauge, the improved gauge fixed action reads \cite{Zwanziger:1992qr,Dudal:2005na}
\begin{equation}
S=S_{YM}+S_{gf}+S_{GZ}\;,\label{fullaction}
\end{equation}
where
\begin{eqnarray}
S_{gf}&=&\int\left(b_A\mathrm{d}\ast Y^A+\overline{c}_A\mathcal{M}{c^A}\right)\;,\nonumber\\
S_{GZ}&=&\gamma^4\kappa^2\int \left[f_{ABC}f^{CDE}Y^A(\ast\mathcal{M}^{-1})^{BD}\ast Y_E+\ast\frac{\mathcal{N}}{\kappa^2}\right]\;.\label{hor1}
\end{eqnarray}
The action $S_{gf}$ is the Landau gauge fixing term supplemented by the Faddeev-Popov term and $S_{GZ}$ is the Gribov-Zwanziger term. The object $\mathcal{M}=\mathrm{d}\ast D$ is the 4-form Faddeev-Popov operator, the fields $c^A$ and $\overline{c}^A$ are the Faddeev-Popov ghost and anti-ghost fields while $b^A$ is the Lautrup-Nakanishi field. The constant $\mathcal{N}=d(N^2-1)$ depends on the spacetime dimension $d=4$ and group dimension $(N^2-1)=24$. The first term in $S_{GZ}$, being non-local, characterizes a highly complicated behavior of the theory. However, this term can be easily localized through the introduction of extra auxiliary fields, see \cite{Zwanziger:1992qr,Dudal:2005na}. Remarkably, the Gribov-Zwanziger term does not spoil the renormalizability of the theory, in fact, the ultraviolet sector remains unchanged and no extra divergences are introduced \cite{Zwanziger:1992qr,Maggiore:1993wq,Dudal:2005na}.

It is easy to check \cite{Baulieu:2008fy,Baulieu:2009xr} that the Gribov-Zwanziger term is the term that breaks BRST symmetry in a soft manner. However, the Gribov parameter tends to vanish at the ultraviolet limit (see Fig.~\ref{fig1}). Thus, at this limit, the BRST symmetry is asymptotically restored at high energies. The consequence for the present gravity model is that, at high energies, the massless theory is consistent at quantum level. Then, at some lower scale, soft BRST breaking takes place. At this scale, the propagators of the fundamental fields cannot be associated with physical excitations anymore \cite{Zwanziger:1992qr,Dudal:2005na,Dudal:2011gd} because of the appearance of complex poles\footnote{The rescaling \eqref{resc1} was employed.}:
\begin{eqnarray} \label{eq:prop}
\left<A^{ab}_\mu A^{cd}_\nu\right>_p &=&\frac{\kappa^2}{2}\left(\delta^{ac}\delta^{bd}-\delta^{ad}\delta^{bc}\right)\left(\frac{p^2}{p^4+\gamma^4}\right)T_{\mu\nu}\;, \nonumber\\
\left<M^{ab}_\mu M^{cd}_\nu\right>_p &=&\frac{\kappa^2}{2}\left(\delta^{ac}\delta^{bd}+\delta^{ad}\delta^{bc}\right)\left(\frac{p^2}{p^4+\gamma^4}\right)T_{\mu\nu}\;, \nonumber\\
\left< \theta^{a}_\mu\pi^{b}_\nu\right>_p&=& \delta^{ab}\frac{\kappa^2}{\gamma^2}\left(\frac{p^2}{p^4+\gamma^4}\right)T_{\mu\nu}\;,
\end{eqnarray}
where $T_{\mu\nu}=\delta_{\mu\nu}-p_\mu p_\nu/p^2$ is the transversal projector in momentum space. The fact that these propagators have no K\"all\'en-Lehmann representation, establishes that these states are removed from the physical spectrum of the theory. This is enough to motivate us to look for new gauge invariant operators to be identified with physical states. As discussed before, we identify these operators with an effective geometry of the spacetime.

It is worth mention that the Gribov-Zwanziger approach can be improved by considering dimension-2 local composite operators and their respective condensates \cite{Dudal:2005na,Dudal:2011gd}. This implementation is known as the \emph{refined Gribov-Zwanziger formalism} and the effect is to deform the horizon function with extra mass parameters associated with the these dimension-2 condensates \cite{Dudal:2011gd}. For the sake of simplicity, we have only considered here the Gribov parameter, which appears to accommodate the relevant effects of this work. Eventually, these condensates might be relevant to adjust the explicit numbers predicted by the theory.

\subsection{Running parameters and one-loop estimates}

\begin{figure}[htb]
	\centering
		\includegraphics[width=0.48\textwidth]{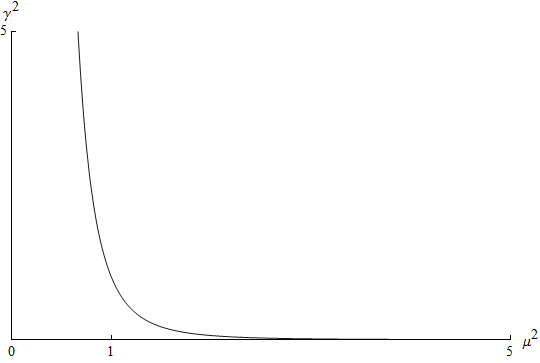}
		\caption{Gribov parameter as function of energy scale. The energy is in units of $\overline{\Lambda}$ and the Gribov parameter in units of $\frac{e^{5/6}}{\sqrt{5}}\overline{\Lambda}^2$.}\label{fig1}
\end{figure}

In perturbative quantum field theory, the running of the parameters can be determined by the renormalization factors of fields and parameters. In the case of the Gribov parameter, it is determined by a gap equation obtained from the minimization of the quantum action with respect to $\gamma^2$. At one-loop, the gap equation is\footnote{It is important to notice that, because we are dealing with Yang-Mills theories, the following computations can be derived in the same manner that of \cite{Dudal:2005na}, only keeping in mind that the fundamental Casimir here is $5$ and the group dimension is $24$.} \cite{Gribov:1977wm,Dudal:2005na}
\begin{equation}
\frac{3N\kappa^2}{4}\int\frac{d^4p}{(2\pi)^4}\frac{1}{p^4+N\gamma^4}=1\;.\label{gap1}
\end{equation}
where $N=5$ and $p$ is the internal momentum. Thus, following \cite{Dudal:2005na}, by employing the $\overline{\mathrm{MS}}$ renormalization scheme at the gap equation \eqref{gap1}, we find
\begin{equation}
\frac{N\kappa^2}{16\pi^2}\left[\frac{5}{8}-\frac{3}{8}\ln\left(\frac{N\gamma^4}{\mu^4}\right)\right]=1\;,\label{gap1a}
\end{equation}
where $\mu$ is the energy scale and the trivial solution $\gamma^2=0$ was excluded. Equation \eqref{gap1a} provides
\begin{equation}
\gamma^4=\frac{e^{5/3}}{N}\mu^4e^{-\frac{8}{3}\frac{16\pi^2}{N\kappa^2}}\;.\label{gap2}
\end{equation}
Recalling that \cite{Gross:1973id,Politzer:1973fx}
\begin{equation}
\frac{N\kappa^2}{16\pi^2}=\frac{1}{\frac{11}{3}\ln\frac{\mu^2}{\overline{\Lambda}^2}}\;,\label{kappa1}
\end{equation}
where $\overline{\Lambda}$ is the renormalization group cutoff, we find (see Fig.~\ref{fig1})
\begin{equation}
\gamma^2=\frac{e^{5/6}}{\sqrt{5}}\overline{\Lambda}^2\left(\frac{\mu}{\overline{\Lambda}}\right)^{-70/9}\;.\label{gamma1}
\end{equation}
We can see from \eqref{gamma1} and Fig.~\ref{fig1} that, as higher the energy, as smaller is the Gribov parameter. This is the expected behavior \cite{Gribov:1977wm,Dudal:2005na,Dudal:2011gd} of $\gamma^2$. At the deep infrared region, however, it seems to diverge. This behavior is an evidence that the semi-perturbative approximation needs improvements. In fact, it is known from lattice simulations for unitary groups that the coupling parameter is finite at the origin \cite{Maas:2011se,Cucchieri:2006xi}. This kind of improvement could affect the infrared sector of the Gribov parameter in such a way that it could also be finite.

\begin{figure}[htb]
	\centering
		\includegraphics[width=0.48\textwidth]{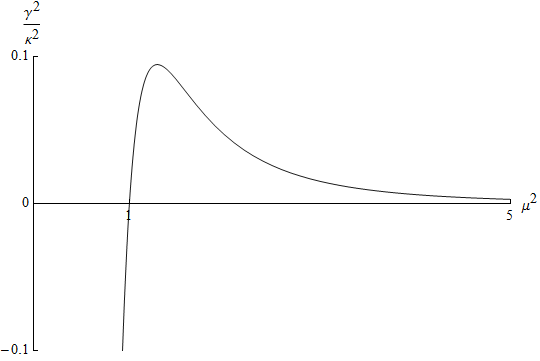}
		\caption{The ratio $r$ as function of energy scale. The energy is in units of $\overline{\Lambda}$ and the ratio $r$ in units of $\frac{55}{48\pi^2}\left(\frac{e^{5/6}}{\sqrt{5}}\right)\overline{\Lambda}^2 $.}\label{fig2}
\end{figure}

Now, the ratio $r=\gamma^2/\kappa^2$ is easily determined from \eqref{kappa1} and \eqref{gamma1}
\begin{equation}
r=\frac{55}{48\pi^2}\left(\frac{e^{5/6}}{\sqrt{5}}\right)\overline{\Lambda}^2\left(\frac{\mu}{\overline{\Lambda}}\right)^{-70/9}\ln\left(\frac{\mu^2}{\overline{\Lambda}^2}\right)\;.\label{ratio1}
\end{equation}
The behavior of $r$ is plotted in Fig.~\ref{fig2}. It is clear that the ratio $r$ has a very interesting behavior. As expected, at high energies, $r$ asymptotically vanishes. Then, in a scale right above $\overline{\Lambda}$, $r$ achieves a maximum. At this region, BRST soft breaking takes place and the rescaling of the fields \eqref{resc1} is allowed. After that, it drops fast to zero at $\mu=\overline{\Lambda}$. Is exactly at this point that the theory suffers the In\"on\"u-Wigner deformation which induces the breaking to the $SO(4)$ theory and the geometric phase starts over. This point is also recognized as the point of phase transition in non-Abelian gauge theories. Below this point, another theory takes place. In the case of quantum chromodynamics, $\overline{\Lambda}_{QCD}\approx 237\mathrm{MeV}$, the Yang-Mills action should be replaced by an action based on hadrons and glueballs excitations. In the case of gravity, the Yang-Mills action must be substituted by a geometrodynamical action.

\begin{figure}[htb]
	\centering
		\includegraphics[width=0.48\textwidth]{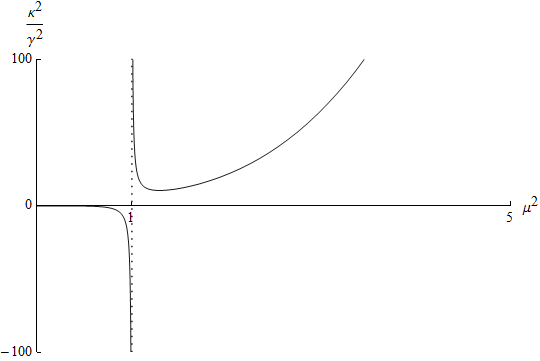}
		\caption{The ratio $r^{-1}$ as function of energy scale. The energy is in units of $\overline{\Lambda}$ and the ratio $r^{-1}$ in units of $\left[\frac{55}{48\pi^2}\left(\frac{e^{5/6}}{\sqrt{5}}\right)\overline{\Lambda}^2\right]^{-1}$.}\label{fig3}
\end{figure}

Below the transition point $\overline{\Lambda}$, where the coupling parameter diverges, the squared coupling parameter acquires negative values. This indicates that below $\overline{\Lambda}$ the perturbative predictions are actually meaningless. However, from lattice predictions \cite{Cucchieri:2009zt}, there are strong evidences that the non-perturbative coupling is actually finite at the origin and presents no divergence at the transition scale. On the other hand, we can argue that the divergence is a strong signal that there is a phase transition at that point. Thus, neither way, the coupling parameter should drop out in favour of an effective coupling. In the case of QCD, it is not known how to obtain this parameter from the dynamics of Yang-Mills theories. However, in the present case, we can interpret the ratio $r$ as the effective coupling (which is related to Newton`s constant). Thus, accepting that for a moment, we can plot the inverse of $r\propto G^{-1}$ (see Fig.~\ref{fig3}) and argue that the graph (and also Fig.~\ref{fig2}) is valid for all scales (except perhaps at the the deep infrared). In this case, after the phase transition (the phase transition is quite evident through a discontinuity at $\mu=\overline{\Lambda}$), the Newton parameter is very large and drops fast to a very small value, as expected. Another interesting point is that, for scales below the phase transition, the ratio $r$ becomes negative. Obviously, this does not mean that the Newton constant is negative. It simply means that a global minus sign will appear at the action \eqref{ym-map-grav}.

The same analysis can be done for the cosmological constant, except that $\Lambda\sim \gamma^2$ and thus, it should be very large. This is actually a very welcome feature because it may compensate the QFT predictions and, eventually, it can provide a small effective value \cite{Shapiro:2009dh}.

Obviously, the present results are obtained by a very naive extrapolation of 1-loop results and must be improved and checked. However, the present extrapolation seems to provide very good qualitative results.

\subsection{Estimates}

The scales $\mu$ and $\overline{\Lambda}$ can be estimated by fixing the current value of $G^{-1}\approx1.491\times10^{32}TeV^2$. Thus, combining \eqref{eq:newton-cosmol} and \eqref{ratio1}, one easily achieves
\begin{equation}
9.932\times10^{31}TeV^2=\overline{\Lambda}^2\left(\frac{\mu}{\overline{\Lambda}}\right)^{-70/9}\ln\left(\frac{\mu^2}{\overline{\Lambda}^2}\right)\;.\label{ratio2}
\end{equation}
We have relative freedom to choose $\mu$ as long as $N\kappa^2/16\pi^2<1$, which means that $\mu^2/\overline{\Lambda}^2> e^{3/11}\approx1.314$ (see \eqref{kappa1}). Let us work at $\mu^2=2\overline{\Lambda}^2$. Thus, $N\kappa^2/16\pi^2\approx 0.393$. This provides $\ln(\mu^2/\overline{\Lambda}^2)\approx 0.693$. Accepting that as reasonable values, we achieve for the renormalization group scale
\begin{equation}
\overline{\Lambda}^2\approx 2.123\times10^{33} TeV^2\;.
\end{equation}
This value provides a time scale of the order of $\tau\sim 10^{-44}s$ which is right below Planck time. Although some approximations and extrapolations have been considered, we can interpret these values as a good result, indicating that the geometric phase of gravity appeared right before Planck scale. This means that right above Planck scale, where quantum mechanics starts to make sense, gravity is already in its geometric phase.

We can also estimate the value of the cosmological constant for the chosen scales. From \eqref{eq:newton-cosmol} and \eqref{gamma1}, it provides
\begin{equation}\label{lambda1loop}
\Lambda^2\approx 1.106\times 10^{32} TeV^2\;.
\end{equation}
This is a huge amount of energy and is three orders of magnitude greater than quantum field predictions \cite{Weinberg:1988cp} of $\Lambda^2_{qft}\sim-3.71\times10^{28}TeV^2$ and would not cancel it to provide the observational data value $\Lambda^2_{obs}\sim1.686\times10^{-68}eV^2$. On the other hand, from electroweak theory \cite{Weinberg:1988cp,Beringer:1900zz} we have $\Lambda^2_{qft}\sim-1TeV^2$, which makes the result even worse. However, for 1-loop approximation at zero temperature, it is quite remarkable that a solution accommodating Newton constant and Planck time could be found. 

\subsection{Consistency of the mapping}

The gravity action \eqref{ym-map-grav} is formally obtained from a consistent map between a gauge theory in a Euclidean space and a gravity theory in a deformed nontrivial spacetime. This map can be described by a bundle map, where the original gauge theory is a principal bundle with structure group $SO(4)$ and base space $\mathbb{R}^4$, and the target principal bundle is a coframe bundle with base space given by $\mathbb{M}^4$ and local $SO(1,3)$ isometry. The consistency of this mapping was analysed in \cite{Sobreiro:2011hb} for the case $SO(m,n)$ with $m+n=5$. The same technique applies here. Essentially, a point $x\in\mathbb{R}^4$ is mapped into a point $X\in\mathbb{M}^4$ where $\mathbb{M}^4$ is the deformed spacetime. To avoid ambiguities between fibers, the map $\mathbb{R}^4\mapsto\mathbb{M}^4$ must be an isomorphism. Then, the structure group is identified with the local isometries of $\mathbb{M}^4$ through $SO(4)\mapsto SO(1,3)$. Moreover, the space of $p$-forms belonging to the cotangent space $T_x^*(\mathbb{R}^4)$ is mapped into the space of $p$-forms in $T^*_X(\mathbb{M}^4)$ with the specific identification given by \eqref{geom2}. The transformation matrix is straightforward computable once the field equations \eqref{eq:eqfield-e}-\eqref{eq:eqfield-m} are solved. In general, the transformation matrix is given by \cite{Sobreiro:2011hb}
\begin{equation}
{L^\nu}_\mu=\left(\frac{\widetilde{g}}{g}\right)^{1/2d}\widetilde{g}^{\nu\alpha} g_{\alpha\mu}\;,\label{map-matrix}
\end{equation}
where $g_{\mu\nu}$ is the metric tensor of the original space and $\widetilde{g}_{\mu\nu}$ is the effective metric tensor. The respective metric determinants $g$ and $\widetilde{g}$ are assumed to be nonvanishing and $d$ is the spacetime dimension. In the present case, $d=4$, $g_{\mu\nu}=\delta_{\mu\nu}$ and $\widetilde{g}_{\mu\nu}$ is a metric tensor describing the de Sitter metric spacetime.

Further, we highlight there is a map from a flat gauge theory onto a Lorentzian gravity theory. We cast the main reasons as follows.
\begin{itemize}
\item[I.] Our starting point, which is a four-dimensional Euclidean gauge theory could perfectly be replaced by a Lorentzian one. In a sense, we could do all previous computations stating that our gauge theory is nothing else but pure Yang-Mills in four Lorentzian dimensions for the gauge group $SL(5,\mathbb{R})$. However, it is well-known that, to do concrete computations in standard Quantum Field Theory, a Wick rotation must be performed in order to have a tractable path integral. At the perturbative level, this is nothing else but a change of variables which can always be done. On the other hand, at the non-perturbative level, the issue of Wick rotating the theory is much more subtle, meaning that its validity is not fully understood. However, we must do computations anyhow and a very common procedure is to start with a theory defined in Euclidean space. Since our symmetry breaking mechanism is based on the existence of a dynamical mass parameter generated by the existence of Gribov copies, which is a genuine non-perturbative phenomenon, we start directly with a gauge theory defined in Euclidean space. Formally, however, we could say that this parameter is there even if we started with a Lorentzian signature, but is clear from the Gribov problem literature, that everything is well-defined in Euclidean space. Therefore, we have two options at this level: \emph{(i)} We state all the well established properties concerning the Gribov parameter in Euclidean space and assume we can Wick rotate at the non-perturbative level which implies a mapping onto a Lorentzian gravity theory; \emph{(ii)} We assume that we could treat the Gribov problem in Lorentzian signature from the very beginning and never speak about Euclidean signature. Anyway, at the end of the story, the signature ``issue" seems to live on the gauge theory side, where we have to assume (or not) that we are free to perform the Wick rotation.

\item[II.] The gravity theory we obtain after the mapping is, in the interpretation of our framework, a \textit{classical} theory. In this regime, we just have to take care of actions and equations of motion, but not of path integrals and quantization devices. Therefore, it seems that the issue of treating a manifold with Euclidean signature or with Lorentzian signature is just a matter of supporting the theory with a causal structure or not. But, at this level, we do not have the same obstructions we do when we are dealing with strongly coupled quantum field theories, and a Wick rotation is supposed to be always possible.

\item[III.] Additionally, we must observe that Eq.~\eqref{map-matrix} is a general mapping which relates the components of a p-form in a manifold $M^d$ with metric $g_{\mu\nu}$ to the components of the corresponding p-form in a target manifold $\widetilde{M}^d$ with metric $\widetilde{g}_{\mu\nu}$. The demonstration is independent of the signatures of the metrics. Moreover, we emphasize that $\tilde{g}=|\det\tilde{g}_{\mu\nu}|$ and $g=|\det g_{\mu\nu}|$. $g$ and $\tilde{g}$ are the absolute values of the $g_{\mu\nu}$ and $\tilde{g}_{\mu\nu}$ determinants, respectively. For completeness, let us provide the following explicit proof.

Let $\mathcal{W}_{\omega\rho}$ to be a Wick transformation defined by
\begin{equation}\label{wick-rot-matrix}
\mathcal{W}_{\omega}^{\phantom{\omega}\rho}=\left(\begin{array}{cccc}
i \\
 & 1 & \makebox(0,0){\text{\huge0}} & \\
 & \makebox(0,0){\text{\huge0}} & \ddots &  \\
 &  &  & 1 \\
\end{array}
\right)~.
\end{equation}
Further, $\mathcal{W}_\omega^{\phantom{\omega}\lambda}=\mathcal{W}^\lambda_{\phantom{\lambda}\omega}$.   and $\mathcal{W}_{\mu\rho}\mathcal{W}^{\rho\nu}=\delta_\mu^{\phantom{\mu}\nu}$. The Wick transformation is achieved by
\begin{equation}\label{wick-transform}
\mathcal{W}_\omega^{\phantom{\omega}\rho}
\widetilde{g}_{\rho\sigma}
\mathcal{W}_\lambda^{\phantom{\lambda}\sigma}=
\overline{\widetilde{g}}_{\omega\lambda}~.
\end{equation}

Now, let us start with $L^\rho_{\phantom{\rho}\omega}=a\tilde{g}^{\rho\lambda}g_{\lambda\omega}$ with $a=\left(\tilde{g}/\overline{g}\right)^\frac{1}{2d}$, where $\tilde{g}=|\det\tilde{g}^{\rho\lambda}|$ and $g=|\det g_{\lambda\omega}|$, while $g_{\lambda\omega}$ and $\tilde{g}^{\rho\lambda}$ are, respectively, the metric of the initial space and the metric of the final curved space. Hence, a Wick transformation can be accounted by $L$ by means of the direct metric mapping:
\begin{equation}\label{wick-metric-map}
\widetilde{g}_{\mu\nu}=L_\mu^{\phantom{\mu}\alpha}L_\nu^{\phantom{\nu}\beta}g_{\alpha\beta}\;.
\end{equation}
Thus, using Eq.~\eqref{wick-metric-map} in Eq.~\eqref{wick-transform}, we obtain
\begin{eqnarray}\label{wick-rot-final}
\overline{\widetilde{g}}_{\mu\nu}&=&\mathcal{W}_\mu^{\phantom{\mu}\alpha}\mathcal{W}_\nu^{\phantom{\nu}\beta}L_\alpha^{\phantom{\alpha}\gamma}L_\beta^{\phantom{\beta}\delta}g_{\gamma\delta}\nonumber\\
\overline{\widetilde{g}}_{\mu\nu}&=&(\mathcal{W}_\mu^{\phantom{\mu}\alpha}L_\alpha^{\phantom{\alpha}\gamma})(\mathcal{W}_\nu^{\phantom{\nu}\beta}L_\beta^{\phantom{\beta}\delta})g_{\gamma\delta}
\end{eqnarray}
Thus,
\begin{equation}\label{wick-rot-L}
\overline{L}_\mu^{\phantom{\mu}\gamma}=\mathcal{W}_\mu^{\phantom{\mu}\alpha}L_\alpha^{\phantom{\alpha}\gamma}
\end{equation}
incorporates the Wick rotation. Moreover $\det{\overline{\widetilde{g}}_{\omega\lambda}}=-\det{\widetilde{g}_{\omega\lambda}}$, hence $\overline{\widetilde{g}}=\widetilde{g}$. Aftermath, we achieve the expected Wick-rotated metric $\overline{\tilde{g}}^{\mu\xi}$ to the final target space.
\end{itemize}

\section{Conclusions}\label{FINAL}

We presented a model where, at high energies, quantum gravity is described by a pure $SL(5,\mathbb{R})$ Yang-Mills action in four Euclidean dimensions while the low energy theory is a geometrodynamical theory containing the Einstein-Hilbert term, cosmological constant and a matter field. The transition between the two phases is mediated by a soft BRST symmetry breaking associated with the Gribov problem. It was found that, at the Landau gauge in one-loop approximation, the ratio between the Gribov parameter and the coupling parameter can be associated with a kind of order parameter that describes the phase transition between both sectors (see Fig.~\ref{fig2}). First, the ratio increases and the soft BRST breaking gradually appears. Then, after a maximum, the ratio goes to zero very rapidly. At this point, the original gauge group suffers a symmetry breaking to the group $SO(4)$ and the geometric phase starts over. The degrees of freedom of the original theory are identified with the vierbein, the spin-connection and a matter field.
 The consequence is that an effective geometry appears. After that, the ratio becomes negative and rapidly increases in magnitude which is good because its inverse is related to Newton's constant (se expression \eqref{eq:newton-cosmol}).
 It is important to keep in mind that this last sector is obtained from an extrapolation that goes beyond perturbation theory.
 However, after the Landau pole, a phase transition is expected and the coupling parameter is meaningful.
 After the phase transition, a different coupling takes place as well as a different theory.
 We have assumed however that the ratio $r$ is a good candidate to describe the phase transition and that it could describe the theory at any scale.
 
Let us make a small interlude about our interpretation on metric free approach. The initial action is an Yang-Mills action in a Euclidean (flat) 4-dimensional spacetime without mass parameters. The mass parameter is generated dynamically due to the Gribov's problem. Only after this dynamical effect we can identify our model with a gravity theory. Our model is developed quite differently from that presented in \cite{Mielke:2012fe} where the authors start from a topological action, which is constructed over a metric-free spacetime. Then, a nontopological theory emerges from a spontaneous symmetry breaking caused by a Higgs-like mechanism. After this breaking, they have reached an emergent gravity model and an effective metric arises.
 
As described in expression \eqref{eq:newton-cosmol}, Newton and cosmological constants are related to the Gribov and coupling parameters. The running of the Newton constant can be visualised at Fig.~\ref{fig3}, which, however, is related to gravity only below $\mu=\overline{\Lambda}$. Moreover, with relatively small expansion parameter and logarithms, we were able to find a solution that accounts for the current value of Newton's constant and predicts a scale of the phase transition right below the Planck scale, as expected. This means that, near to the Planck scale, gravity would already be at the geometric phase. However, the predicted value for the cosmological constant does not account for the cancellation of quantum field theory predictions. Nevertheless, we found remarkable that these results could be obtained from 1-loop approximations. It is also remarkable that the same mechanism that describes confinement in QCD can be employed to generate gravity. Improved explicit computations are required indeed, but we leave it for future investigation.

Let us take a look at the classical theory described by the action \eqref{ym-map-grav}. The Newton constant is a global factor and will not affect our analysis. The first term of this gravity action is a squared curvature term and is proportional to $\Lambda^{-2}$. Since the cosmological constant is very large, this term is negligible when compared, for instance, to the Einstein-Hilbert term. Thus, forgetting about the matter field, the only difference between the action \eqref{ym-map-grav} and general relativity with cosmological constant is the squared torsion term. However, torsion is expected to be very small and account only for a suitable coupling with fermions.

Concerning the matter field, except for the mass terms (the last two terms), the matter action is also proportional to $\Lambda^{-2}$. The first and third terms at the second line of \eqref{ym-map-grav} tells us that this field interacts with gravity. On the other hand, looking at the mass terms (see also the field equations \eqref{eq:eqfield-m}), we can see that the mass is proportional $\Lambda^2$. Thus, since this is a huge amount of mass, the matter field has a very small range. In fact, the field $m$ only interacts with gravity and has a very small range, this makes the matter field a potential candidate for dark matter. In the future, we intent to exploit these issues as well.

About the classical limit of the equation \eqref{eq:eqfield-e} we stress that, under small curvatures and due the presence of the high-valued renormalized cosmological constant displayed in \eqref{lambda1loop}, the usual GR limit is attained. Let us clarify this point looking at \eqref{eq:eqfield-o}. If $(3/\Lambda^2)D\star R^{ab}\approx 0$ (for small curvatures), then we obtain an algebraic equation for the torsion which has $T=0$ as a possible solution. Under such limits, $i.e.$, small curvatures and $T=0$, we have $(3/2\Lambda^2)R^{\mathfrak{bc}} \star (R_{\mathfrak{bc}}e_{\mathfrak{a}})\approx 0$ and the vacuum Einstein's field equations with cosmological constant are recovered as well for the $m=0$ solution, otherwise, we have GR coupled to a massive matter field and cosmological constant. Nevertheless, we also have freedom to insert a non-vanishng torsion scenario as well. However, such investigations on nonvanishing torsion and/or high curvature were left to future investigations.

Three last comments are in order. The first one concerns the Weinberg-Witten theorems \cite{Weinberg:1980kq}. It was discussed in \cite{Sobreiro:2012dp,Sobreiro:2012iv} the fact that the present model does not violate these theorems. Essentially, these theorems forbid: (i) \emph{massless charged states with helicity $j>1/2$ which have a conserved Lorentz-covariant current} and (ii) \emph{massless states with helicity $j>1$ which have conserved Lorentz-covariant energy-momentum tensor}. First of all, the theory has several mass gaps. In here, only the Gribov parameter was considered. Nevertheless, it was enough to provide a nontrivial pole for the propagators of the theory. Second, and perhaps more important, we have not constructed spin-2 states from composite operators. We have identified some fields (see expression \eqref{geom2}) and composite operators (see expression \eqref{geom1}) with geometric quantities of a different spacetime. In fact, Weinberg-Witten theorems depend on Lorentz global symmetry, which is lost during the mapping.

The second comment concerns the BRST symmetry breaking mechanism in our model, which is part of the geometrodynamical phase ascending. There is another class of gravity theories \cite{Mielke:2007zs} that uses BRST quantization to obtain an emergent gravity as the dominant classical contribution to the path integral.

The third comment is about the graviton itself: where are the spin-2 excitations usually associated with gravitons? The answer is simple, there are not spin-2 excitations. Due to the hierarchy problem, the low energy geometrodynamical theory here occurs at a very high energy scale (near Planck scale) when compared to the other interactions. Only after a considerable decrease of the energy, the geometric theory could be linearized in order to background fluctuations be associated with spin-2 waves. Thus, at quantum level, the present model is made of spin-1 ``gravitons'' described by the gauge field. Then, a kind of attraction/repulsion behaviour can be expected. This is a very good feature because it could prevent the usual singularity problems that appear in general relativity. We can then look at this theory as having three stages: The first stage is beyond Planck scale where gravity is a pure massless Yang-Mills gauge theory described by spin-1 excitations. Then, a second stage takes place when soft BRST symmetry breaking enforces the theory to change to a geometrodynamical theory. This sector is a low energy limit. However, it is at high energies when compared to the other fundamental interactions. Then, a third stage can be originated for weak curvature regime. At this last stage, the theory can be linearized around a flat background and then spin-2 perturbations arise as in usual linearized gravity.
 
In summary, we have shown that gravity, a potential candidate to dark matter and dark energy can arise as an effective feature from a gauge theory. Although improvements must be performed, this theory provides good estimates for a semi-perturbative one-loop approximation.

\section*{Acknowledgements}

The Conselho Nacional de Desenvolvimento Cient\'{i}fico e Tecnol\'{o}gico (CNPq-Brazil), The Coordena\c c\~ao de Aperfei\c coamento de Pessoal de N\'ivel Superior (CAPES) and the Pr\'o-Reitoria de Pesquisa, P\'os-Gradua\c c\~ao e Inova\c c\~ao (PROPPI-UFF) are acknowledge for financial support.

\end{document}